\documentclass[12pt]{article} 
\usepackage{amssymb}
\usepackage{graphicx}
\begin{document} 
\begin{center}

{\bf Excess of soft dielectrons and photons}

\vspace{2mm}

I.M. Dremin\footnote{e-mail: dremin@lpi.ru}\\

{\it Lebedev Physical Institute, Moscow, Russia}

\end{center}

                                             
\vspace{1mm}

\begin{abstract}
Spectra of unbound electron-positron pairs (dielectrons, in brief)
and photons from decays of
parapositronia produced in ultraperipheral collisions of electrically charged
objects are calculated. Their shapes at energies of the NICA collider are
demonstrated. Soft dielectrons and photons are abundantly produced. 
The relevance of these processes to the astrophysical problem
of cooling electron-positron pairs and the intense emission of 511 keV photons
from the Galactic center is discussed.
\end{abstract}

\vspace{1mm}


\vspace{1mm}

Excess of soft dileptons and photons over their supply from all well established 
sources is a widely known fact in particle physics and in astrophysics.
It is reasonable to assume that electromagnetic fields are responsible for
this effect. The outcome of a collision of two charged objects is especially soft
if they do not come too close to one another but pass at large distances (impact 
parameters) and only their electromagnetic fields interact. These processes 
were named ultraperipheral because of the large spatial extension of 
electromagnetic forces. It is shown here that very soft dielectrons and photons 
are produced in these grazing collisions.

Electromagnetic interactions of rapidly moving charged objects were
first considered by Fermi \cite{fer1, fer2} in 1924. He derived the formula
for the intensity of the electromagnetic field created in such processes.
This field can be treated in terms of equivalent photons as was done by
Weizs\"{a}cker \cite{wei} and Williams \cite{wil} in 1934. The same year,
Landau and Lifshitz \cite{lali} published their result on the energy
dependence of the cross section of dielectrons production in 
high energy cosmic rays ions collisions. This is still the most rapidly
increasing (with energy increase) cross section known in particle physics.
Asymptotically it is proportional to $\ln^3\gamma $, where $\gamma =E_c/M_c$ 
is the Lorentz-factor, $E_c$ and $M_c$ are the energy and the mass of a colliding
particle. The two-photon collisions in electromagnetic fields  
create soft dileptons and $CP$-even resonances (see, e.g., 
\cite{ufnd, bgms, vyzh, bb1, kn, bkn, balt, kgsz, seng, skl, zha, kmr, geom}). 
These processes have been
recorded in collider experiments. Further progress in measurements of ever 
softer products is required. The ultraperipheral processes can be responsible 
for the recorded excess of soft dileptons in accelerator experiments as well as 
for the rapid cooling of the electron-positron clouds in the Universe and 
for the famous excess of 511 keV photons emitted from the Galactic center.

The spectra of dielectrons and photons created in ultraperipheral collisions
can be obtained from the general expression for the total cross section
\begin{equation}
\sigma _{up}(X)=
\int dx_1dx_2\frac {dn}{dx_1}\frac {dn}{dx_2}\sigma _{\gamma \gamma }(X),
\label{e2}
\end{equation}
derived in the equivalent photon approximation. Feynman diagrams of 
ultraperipheral processes contain the subgraphs of two-photon interactions 
leading to production of some final states $X$ (e.g., $e^+e^-$ pairs).
These blobs can be represented by the cross sections of these processes.
Therefore, $\sigma _{\gamma \gamma }(X)$ in (\ref{e2}) denotes the total cross 
section of interaction of two photons from the electromagnetic clouds 
surrounding colliding objects and $dn/dx_i$ describe the density of photons 
carrying the share $x_i$ of the energy of the objects. 

In what follows, we examine the collisions of heavy Au$^{79+}$-ions at the NICA
collider. The photon fluxes $dn/dx_i$ are calculated from the flux of 
electromagnetic fields surrounding the colliding ions. They are shown in many 
textbooks on quantum electrodynamics. The~distribution of equivalent photons 
generated by a moving nucleus with the charge $Ze$ and carrying a fraction of 
the nucleon energy $x$ integrated over the transverse momentum up to some value 
(see, e.g., \cite{blp}) can be denoted as
\begin{equation}
\frac {dn}{dx}=\frac {2Z^2\alpha }{\pi x}\ln \frac {u(Z)}{x}.
\label{flux}
\end{equation}
Soft photons carrying small fractions $x$ of the nucleon energy dominate in 
these fluxes. The physical meaning of the ultraperipherality parameter $u(Z)$ 
is the ratio of the maximum 
adoptable transverse momentum to the nucleon mass as the only massless 
parameter of the problem. It differs numerically in various approaches 
\cite{ufnd, bgms, vyzh, bb1, kn, bkn, balt, kgsz, seng, skl, zha, kmr, geom}. 
It depends on charges $Z_ie$, sizes and impact parameters of colliding objects
(form factors and absorptive factors) as well as, in principle, on the considered 
processes. 

The cross section $\sigma _{\gamma \gamma }(X)$ to be inserted 
in (\ref{e2}) in case of creation of the unbound dielectrons $X$
looks \cite{brwh, blp} as
\begin{equation}
\sigma _{\gamma \gamma }(X)=\frac {2\pi \alpha ^2}{M^2}
[(3-z^4)\ln \frac {1+z}{1-z}-2z(2-z^2)], 
\label{mM}
\end{equation}
where $z=\sqrt {1-\frac {4m^2}{M^2}}$, $m$ is the electron mass and $M$ is the 
dielectron mass. This cross section tends to 0 at the threshold $M=2m$ and 
decreases as $\frac {1}{M^2}\ln M$ at very large $M$.

The distribution of masses $M$ of dielectrons is obtained after inserting
Eqs (\ref{flux}), (\ref{mM}) into (\ref{e2}) and leaving free one integration
there. One gets 
\begin{equation}
\frac {d\sigma }{dM}=\frac {128 (Z\alpha )^4}{3\pi M^3}
[(1+\frac {4m^2}{M^2}-\frac {8m^4}{M^4})
\ln \frac {1+\sqrt {1-\frac {4m^2}{M^2}}}{1-\sqrt {1-\frac {4m^2}{M^2}}}-
(1+\frac {4m^2}{M^2})\sqrt {1-\frac {4m^2}{M^2}}]
\ln ^3\frac {u\sqrt {s_{nn}}}{M}, 
\label{sM}
\end{equation}
where $\sqrt {s_{nn}}$ is the c.m.s. energy per a nucleon pair. The dielectron
distribution (\ref{sM}) is shown in Fig. 1 for three NICA energies ranging 
from 11 GeV to 8 GeV and 6.45 GeV per nucleon. The parameter $u$=0.02 has been
chosen in accordance with its value obtained in Ref. \cite{vyzh} where
careful treatment of nuclei form factors is done. The total cross section of
ultraperipheral production of unbound pairs is
\begin{equation}
\sigma (ZZ(\gamma \gamma )\rightarrow ZZe^+e^-)=\frac {28}{27}
\frac {Z^4\alpha ^4}{\pi m^2}\ln^3\frac {u^2s_{nn}}{4m^2}.
\label{vz}
\end{equation}

\begin{figure}

\centerline{\includegraphics[width=16cm, height=14cm]{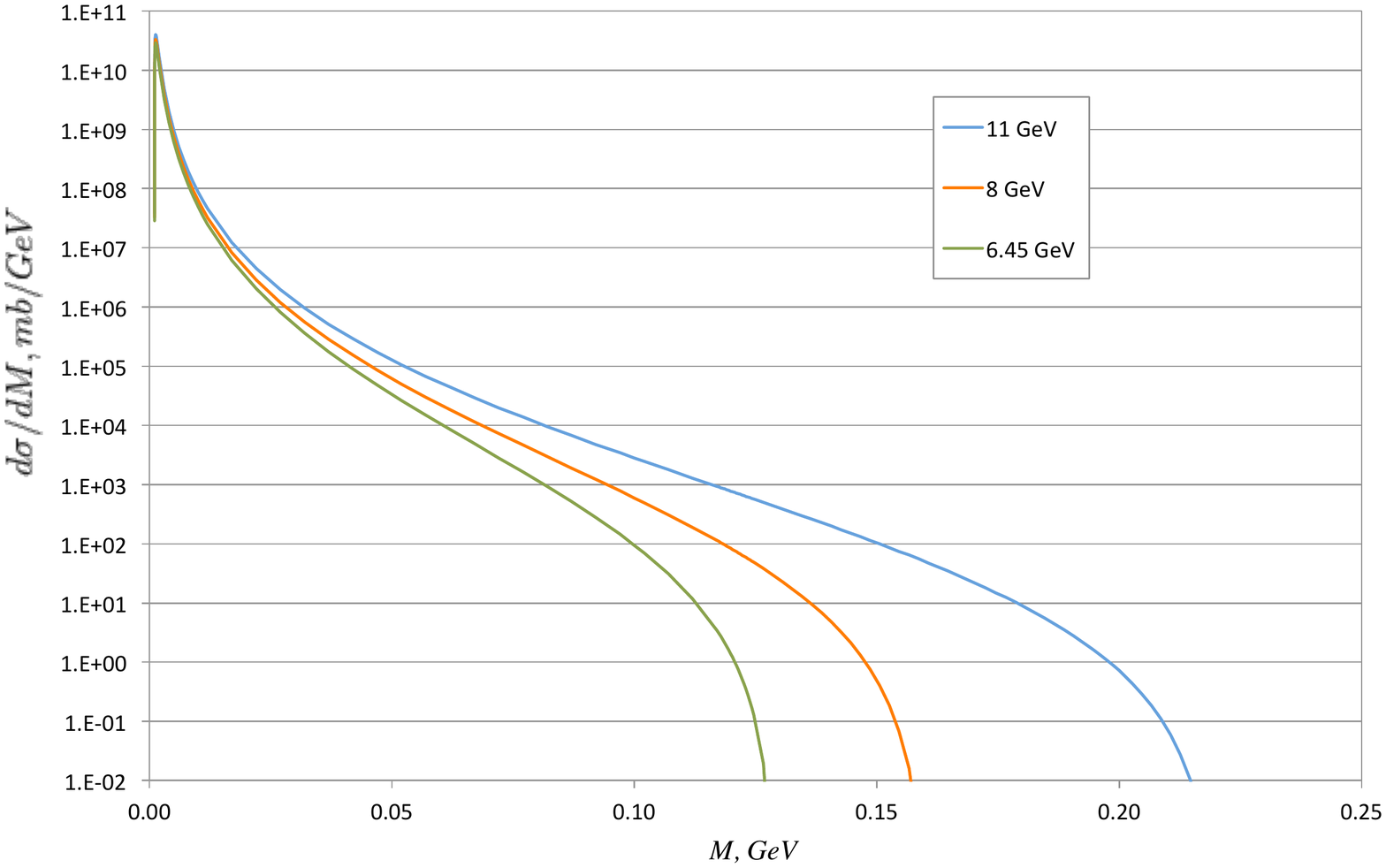}}

Fig. 1. The distribution of masses of dielectrons produced in
ultraperipheral collisions at NICA energies $\sqrt {s_{nn}}$=11 GeV 
(blue, upper), 8 GeV (red, middle), 6.45 GeV (green, lower).
\end{figure}

The sharp peak at very low masses $M$ in Fig. 1 demonstrates the most important
feature of ultraperipheral processes - the abundant production of soft 
dielectrons with masses of the order of several electron masses $m=0.511$ MeV 
(effectively, less than 50 MeV).
Their direct registration is a major problem in accelerator experiments
as shown in a recent paper of ALICE Collaboration
at LHC on observed soft-dielectron excess in proton-proton collisions at 
$\sqrt s$=13 TeV \cite{ali}. Unfortunately the dielectrons with masses less 
than 150 MeV were inaccessible there for recording. However, recording 
dielectrons with masses down to this limit marks substantial progress 
relative to the previosly achieved lower limit, which was several times higher. 
Special adjustment of the magnetic fields at LHC was required to get this value. 
As can be seen in Fig. 1, the limit of 150 MeV is still well above the masses 
in the area of abundant ultraperipheral production of dielectrons.

Surely, the excess of dielectrons might be recorded at NICA energies but 
some special experimental technique is needed. The spectrum of dielectrons 
created per 1 GeV interval shown in Fig. 1 is extremely soft. The experimental 
challenge of dielectron measurements is the high combinatorial background as 
well. 
The background from all other nuclear sources including thermal radiation 
and freeze-out coctail is negligibly small compared to the contribution of
ultraperipheral processes at low masses shown in Fig. 1. That is easily seen
if Fig. 1 is matched with the left panel in figure 17 of Ref. \cite{sen} 
which depicts a dielectron invariant mass spectrum $dN/dM$ simulated for central 
Au+Au collisions at $\sqrt {s_{nn}}$=6.45 GeV \cite{ably}. For comparison
of the ordinates, the values of $d\sigma /dM$ in Fig. 1 must be normalized to
the total cross section about 5 kb at 6.45 GeV. The extremely 
strong excess of dielectrons with masses below 10 MeV is predicted.

Some slight excess of positrons can appear in the observed state of
dileptons due to the so-called bound-free process when the created
electron is captured by one of the colliding nuclei. This effect can limit the
luminosity and lead to some damage of the collider. Its cross section was 
recently estimated at NICA energies in \cite{bks}. It is much smaller than 
that for unbound pairs. Therefore, this effect does not change our
conclusions.

The NICA energy interval also permits the ultraperipheral production
of positronia and $\pi ^0$-mesons. Creation of parapositronia is especially
interesting because their decay products at rest are two gamma-quanta with
energies about 511 keV. For $\pi ^0$'s this energy is 67.5 MeV.
Their recording requires a different technique
and looks somewhat simpler than in the case of dielectrons.
The origin of the strong line at 511 keV 
emitted from the Galactic center has not been explained yet \cite{sieg}. 
Studies at NICA collider can help in resolving this puzzle. 
 
The total cross section of the direct production of parapositronia in
two-photon interactions (\ref{e4}) is much lower than the cross
section for creation of unbound pairs (\ref{vz}): 
\begin{equation}
\sigma_{Ps}=\frac {16Z^4\alpha ^2\Gamma }{3m^3}\ln ^3\frac {u\sqrt {s_{nn}}}{m}.
\label{e4}
\end{equation}
However, the final products differ. 
\begin{figure}

\centerline{\includegraphics[width=16cm, height=14cm]{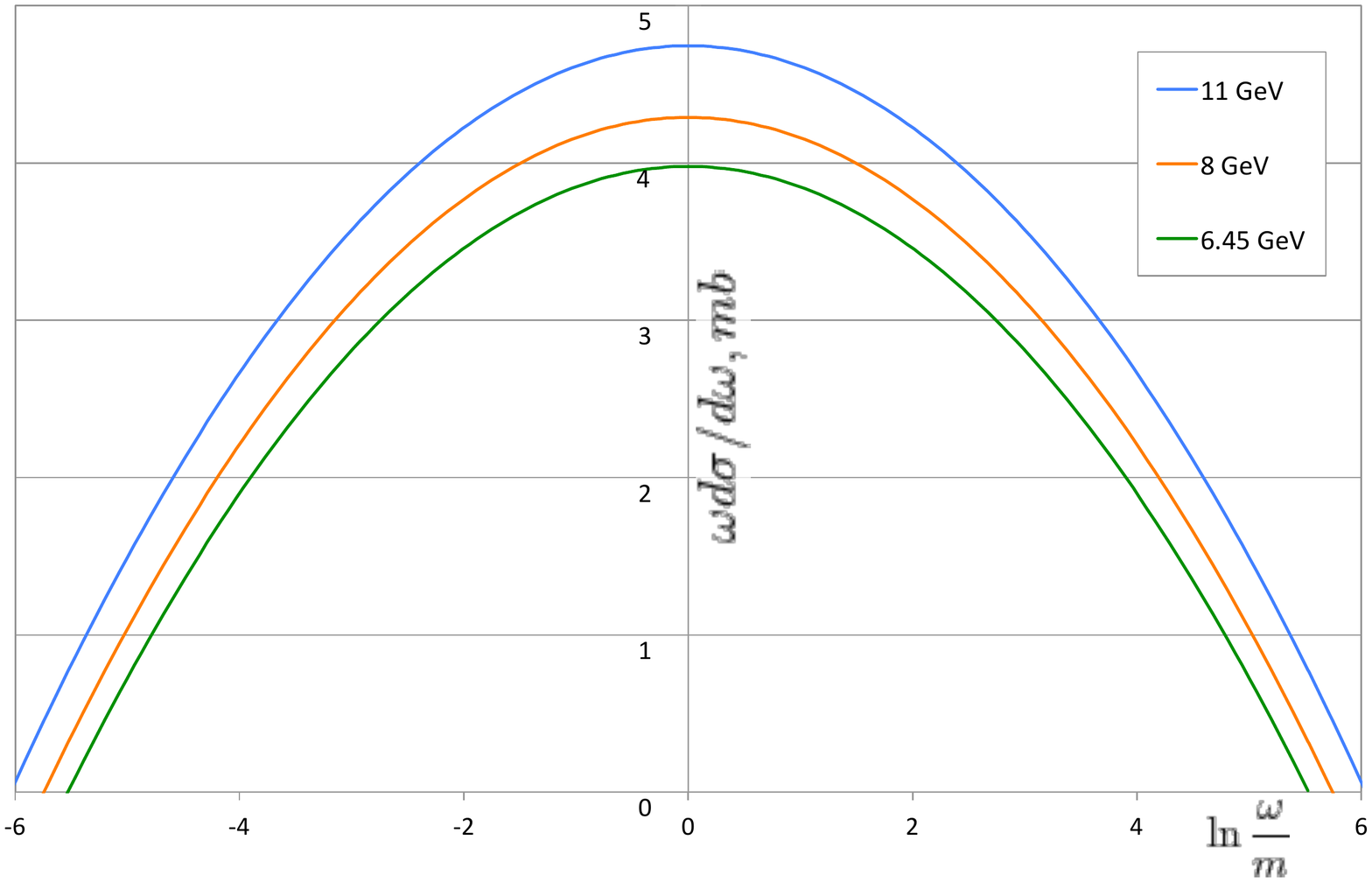}}

Fig. 2. The energy distribution of gamma-quanta from decays of
parapositronia produced in
ultraperipheral collisions at NICA energies $\sqrt {s_{nn}}$=11 GeV 
(blue, upper), 8 GeV (red, middle), 6.45 GeV (green, lower).
\end{figure}
One must record the gamma-quanta instead of electrons.
Their energy distribution is obtained from Eq. (\ref{e2}) by omitting
one of the integrations there and inserting the resonance cross section
for $\sigma _{\gamma \gamma }(X)$ (see, e.g., \cite{ufnd}). For Au-Au 
collisions at NICA one gets
\begin{equation}
\omega \frac {d\sigma }{d\omega }=\frac {4Z^4\alpha ^2\Gamma }{m^3}
(\ln ^2\frac {u\sqrt {s_{nn}}}{m}-\ln^2\frac {\omega }{m}), 
\label{gam}
\end{equation}
where $\Gamma \approx 5.2\cdot 10^{-15}$ GeV is the decay width of the 
parapositronium. 

This distribution (\ref{gam}) is shown in Fig. 2 as 
$\omega d\sigma /d\omega $ for three energies of NICA ranging 
from 11 GeV to 8 GeV and 6.45 GeV per nucleon. Again, the parameter $u$=0.02 is
chosen in accordance with its value obtained in Ref. \cite{vyzh}.

As expected, the photon spectra are concentrated near the electron mass
511 keV and they are rather wide. The motion of parapositronia produced in
ultraperipheral collisions at high enough energies of NICA is responcible for 
the
broadened spectra in Fig. 2. In general, the direct ultraperipheral production 
of parapositronia is about a million times less effective than the creation of
dielectrons as estimated from Eqs (\ref{vz}), (\ref{e4}).

The estimates done above in the equivalent photon approximation are, strictly 
speaking, valid at very high energies. The preasymptotical corrections can
become important at NICA energies (see, e.g., \cite{ufnd, bks}) and ask for
different values of the ultraperipherality parameter $u$. On the other hand,
its application at low lepton masses (electrons!) is also questionable.
Nevertheless, the qualitative feature of soft outcome in these processes
is well established.

Coming now to geophysics, it is worthwhile to remark that the 511 keV photons 
from the parapositronia clouds have been recorded \cite{abc} during the 
thunderstorms in the Earth atmosphere. Strong electromagnetic fields there favor 
somehow the creation of parapositronia at rest. Regarding the astrophysical 
observation of a narrow line at 511$\pm $2.2 keV stemming from the Galactic 
center, it would be reasonable to assume that this line originates from 
decays of very slow parapositronia. Huge clouds of them formed during 
"the Galactic thunderstorms" must exist in the Galactic center to emit this 
strong and narrow line. The initially created soft dielectrons can be in charge
of their formation. Different astrophysical objects are claimed to produce
the dense clouds of the electron-positron plasma. Besides, some microscopic 
mechanism of soft dielectron production active inside the objects must be 
at the origin of these parapositronia in the Universe. 

The process of ultraperipheral production of dielectrons by the electromagnetic 
fields of moving charged particles is a good candidate for providing 
extremely soft dielectrons as shown above. It is especially intense
in collisions of the relativistic heavy ions. Here, the cross sections are 
proportional to the factor $Z^4\alpha ^4$ (i.e., enlarged by $Z^4$ compared
to the singly charged particles) and strongly increase with increasing energy
as $\ln ^3\gamma $. However, their values are still large enough even for
protons with lower energies because of the smaller ultraperipherality parameter.
As shown in Fig. 1, the masses of produced dielectrons are very small. 
It means that the relative velocities of electrons and positrons are not high. 
Dielectrons, produced initially by this process, would further cool down, e.g., 
by bremsstrahlung or
again due to mutual ultraperipheral collisions of (already soft!) electrons 
and/or positrons with creation of additional dielectrons in the reaction
$e^{\pm}e^{\pm}\rightarrow e^{\pm}e^{\pm}e^+e^-$. Thus the density of the 
electron-positron plasma increases while the energies 
of electron-positron pairs diminish so that they come almost to rest and are 
able to form extremely slow moving parapositronia. The cross section of this
process at small dielectron masses is quite large $\sigma \propto \pi r_0^2
\approx 250$ mb where $r_0\approx 2.8\cdot 10^{-15}$ m is the classical 
electron radius. The width of the photon spectrum 
would become much narrower than demonstrated in Fig. 2 which was computed 
for direct ultraperipheral production of parapositronia at rather high energies
of colliding nuclei. The competition between 
the increasing density of the electron-positron plasma and the decreasing cross 
section of ultraperipheral processes near the threshold must be studied. This 
new hypothesis needs further foundations and theoretical efforts to become 
acceptable.

\vspace{6pt}
{\bf Acknowledgments}

{This work was supported by the RFBR project 18-02-40131.

I am indebted to V.G. Terziev for reading 
the manuscript and help with Figures.}

\vspace{6pt}

The author declares no conflicts of interest.


\end{document}